\documentclass[12pt]{article}
\usepackage{amsmath}
\usepackage{graphicx,psfrag,epsf}
\usepackage{enumerate}
\usepackage[dvipsnames]{xcolor}
\usepackage{natbib}
\usepackage{algorithm}
\usepackage{setspace} 
\usepackage{array}
\usepackage{soul}
\usepackage{booktabs}

\usepackage{comment}
\usepackage{algorithmic}
\usepackage{url} % not crucial - just used below for the URL 
\usepackage{float} 

\newcommand{\blind}{1}

\usepackage{graphicx,psfrag,epsf}
\usepackage{caption}
\usepackage{hyperref}
\usepackage{enumerate}
\usepackage{natbib}
\usepackage{url} % not crucial - just used below for the URL 
\usepackage{hhline,array,amsmath,graphicx}
\usepackage{pifont,latexsym,ifthen,rotating,calc}
\usepackage{siunitx}
\usepackage{etoolbox}
\usepackage{mathtools}
\usepackage{amssymb}
\usepackage{amsthm}
\usepackage{bbm}
\usepackage{pdflscape}
\usepackage{rotating}
\usepackage{float}
\usepackage{bm}
\usepackage{enumitem}
\usepackage[mathscr]{euscript}
\usepackage{multirow}
\usepackage{xcolor}

\begin{document}

\def\spacingset#1{\renewcommand{\baselinestretch}%
{#1}\small\normalsize} \spacingset{1}

%%%%%%%%%%%%%%%%%%%%%%%%%%%%%%%%%%%%%%%%%%%%%%%%%%%%%%%%%%%%%%%%%%%%%%%%%%%%%%

\if1\blind
{
  \title{\bf Exploring Causal Mediation Analysis in Bacterial Vaginosis: Challenges, Approaches, and Implications for Women's Health}
  \author{Debarghya Nandi \textsuperscript{1}, 
            Soumya Sahu \textsuperscript{1}, 
            Supriya Mehta \textsuperscript{2},
            Dulal K. Bhaumik \textsuperscript{1}\\
    \textsuperscript{1} Division of Epidemiology and Biostatistics, University of Illinois Chicago\\
     \textsuperscript{2} Division of Infectious Diseases, Rush University}
  \maketitle
} \fi

\if0\blind
{
  \bigskip
  \bigskip
  \bigskip
  \begin{center}
    {\LARGE\bf Exploring Causal Mediation Analysis in Bacterial Vaginosis: Challenges, Approaches, and Implications for Women's Health}
\end{center}
  \medskip
} \fi

\begin{center}
 {\bf{Abstract}}   
\end{center}

Bacterial Vaginosis (BV) affects nearly 23-29\% of women worldwide and increases risk of miscarriage, preterm birth, and sexually transmitted infections. It involves a shift in the vaginal microbiome from Lactobacillus dominance to a diverse bacterial composition. Understanding causal pathways linking behavioral factors to BV risk is essential for effective intervention. Observational studies have identified pathogenic bacteria associated with BV, and causal mediation analysis can clarify how behaviors like sexual activity influence the microbiome. Analyzing microbiome data is complex due to its high-dimensional and compositional nature, often challenging traditional statistical methods, especially with small samples. This article presents various approaches to measure causal mediation effects, emphasizing the benefits of an empirical distribution method for small samples, and outlines models for mediators, exposure, and outcomes, aiming to identify taxa that mediate the exposure-outcome relationship in BV, concluding with a revisit of the motivational example and model identification.\\

\noindent%
\vfill

\addtolength{\textheight}{.2in}%
\newpage
\spacingset{1.9} % DON'T change the spacing!
\section{Introduction}
\label{sec:intro}

Bacterial Vaginosis (BV) is a common reproductive tract infection, affecting nearly 23-29\% of women worldwide \cite{who2025bv}. BV is typically characterized by a shift in the vaginal microbiome, one that is Lactobacillus-dominated state to a more diverse one, containing many different bacteria. In pregnant women, BV significantly increases the risk of miscarriage, preterm birth, and other adverse pregnancy outcomes. Additionally, BV increases the risk of acquiring HIV and other sexually transmitted infections \cite{nava2021bv, cohen2012bv}. Preventing BV is an important public health objective and a priority for clinical interventions. The vaginal microbiome plays a key role in maintaining reproductive health and protecting against pathogenic infections. There is a growing interest in moving beyond correlations to identify the causal pathways through which behavioral exposures may influence BV risk through changes in the vaginal microbiome. Understanding these pathways is essential for developing effective interventions aimed at preventing BV and improving women’s reproductive health.

Prior observational studies identified associations between certain pathogenic bacteria, such as \textit{Gardnerella spp.}, \textit{Prevotella spp.}, etc.   and an increased risk of BV \cite{who2025bv}. Causal mediation analysis may help us understand the complex pathways through which behavioral factors, such as sexual activity, lead to changes in the composition of  vaginal microbiome, potentially resulting in BV \cite{hoffmann_microbiome_2016,mehta_host_2020,mehta_analysis_2023}. Accurate identification of these pathways is needed for developing therapeutic strategies to enhance the health of women affected by BV.

Microbiome data is primarily generated using high-throughput sequencing technologies, such as 16S rRNA gene sequencing or shotgun metagenomic sequencing. This is followed by taxonomic profiling of microbial communities at the genus or species level. Conducting causal mediation analysis in microbiome studies is particularly challenging due to several inherent complexities. Firstly, microbiome data are high-dimensional. Secondly, the data are compositional, meaning the relative abundance of microbial taxa within a sample sums to one or a constant, which complicates the application of traditional statistical methods that assume independence between variables. Third, microbiome data often exhibit high levels of sparsity and heterogeneity, causing traditional asymptotic methods to yield unsatisfactory results, especially for small samples \cite{pan2021}. Our aim is to address all these issues while analyzing microbiome data to ensure we estimate valid causal inference effects.

Existing statistical methods are inadequate for making inferences about causal mediation effects, particularly when non-linear models are used for microbiome data analysis with small samples. This article employs multiple approaches to measure causal mediation effects via confidence intervals and demonstrates that for small samples, one approach based on the empirical distribution has a distinct advantage over the others. However, all approaches perform satisfactorily for large samples.

The rest of the paper is organized as follows. In the next section \ref{sec:motivation}, we describe a motivational application and graphically present the aforementioned data complexity.  In Section \ref{sec:mediator-model}, we present three models; the first one is a probability model for mediators addressing the concerns of zero inflation, varying dispersions and sequencing depths of taxa count. The second model is for a binary exposure with a set of covariates that confound the outcome-exposure relationship. The third model expresses the relationship between the outcome and exposure through causal pathways. In addition, we develop a weight function involving the mediator and exposure models and then use it  to estimate parameters of the outcome-exposure model. In Section \ref{sec:causal_med}, following the concept of natural direct and indirect effects, we estimate the causal mediation effect. In Section \ref{sec:fiducial}, an empirical distribution of the estimate is developed using the concept of fiducial approach and then used it to construct the generalized confidence interval (GCI) for causal mediation effect. In Section \ref{sec:simulation}, via extensive simulations, we provide adequate confidence on identifying taxa pathways through sexual activity potentially leading to acquiring BV. We revisit the motivational example in Section \ref{sec:data_analysis}, and identify  taxa that are causally mediating the exposure-outcome relation. Finally, we conclude the article in Section \ref{sec:discussion}.

\section{Motivational Example}
\label{sec:motivation}
\cite{mehta_analysis_2023} conducted a study(known as the  CaCHe Study) with a cohort of 436 school girls in Kenya, primarily focused on identifying risk factors associated with BV. The study collected data on various parameters, including COVID-related stress, sociodemographic factors, sexual behavior, vaginal microbiome composition, and the BV status of the study subjects. The primary objective of this article is to determine whether engaging in sex with multiple sex partners (\textit{sp6mosm}) increases the risk of acquiring BV and to identify specific taxa that mediate this process.
\begin{figure}[H]
\begin{flushleft}
    \includegraphics[width=1\textwidth, height = 0.30\textheight]{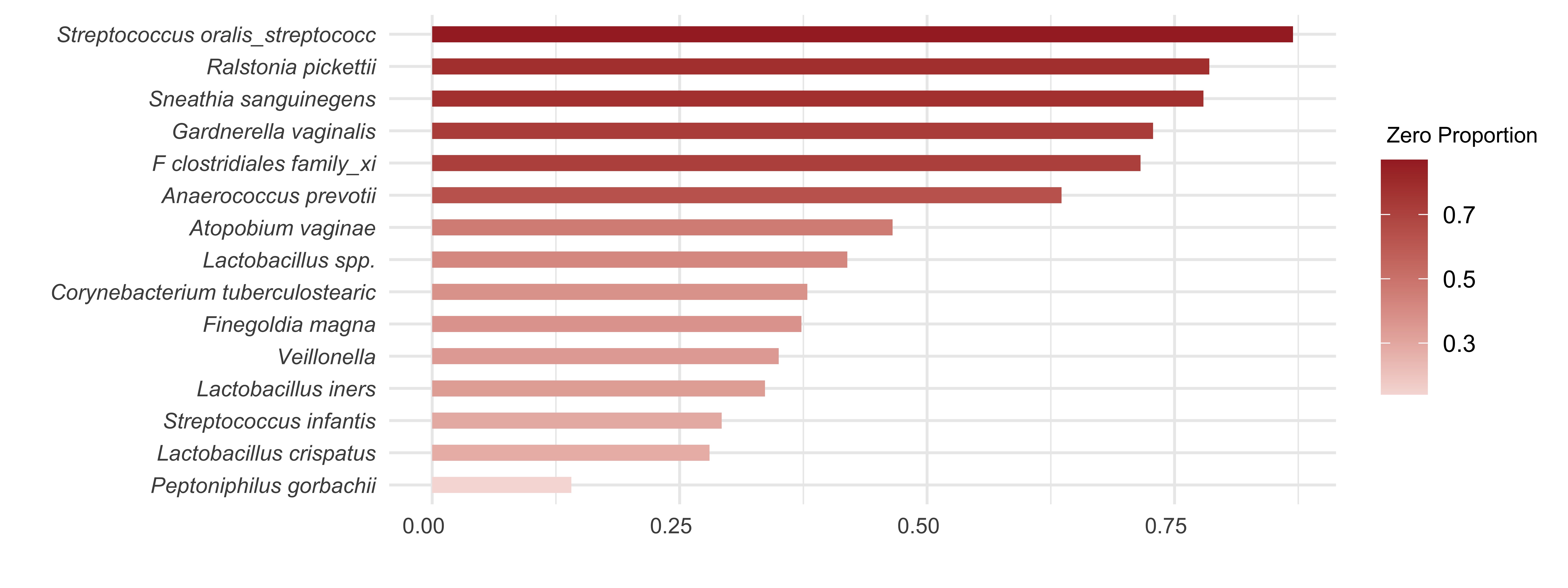}
    \end{flushleft}
        \caption{ Proportion of zeros in the 15 taxa accounting for the highest number of sequence reads in the overall sample.}
        \label{fig:pi_plot}
\end{figure}

The study data features most of the complexities as stated in the introduction. The bar chart in Figure \ref{fig:pi_plot} highlights how the proportions of zeros vary across multiple taxa. For instance, \textit{Peptoniphilus gorbachii} has a low sparsity, with just 10$\%$ zeros, while both \textit{Gardnerella vaginalis} and \textit{Sneathia sanguinegens} exhibit high sparsity, more than 70$\%$.

\begin{figure}[H]
    \hfill
        \includegraphics[width=\textwidth, height = 0.4\textheight]{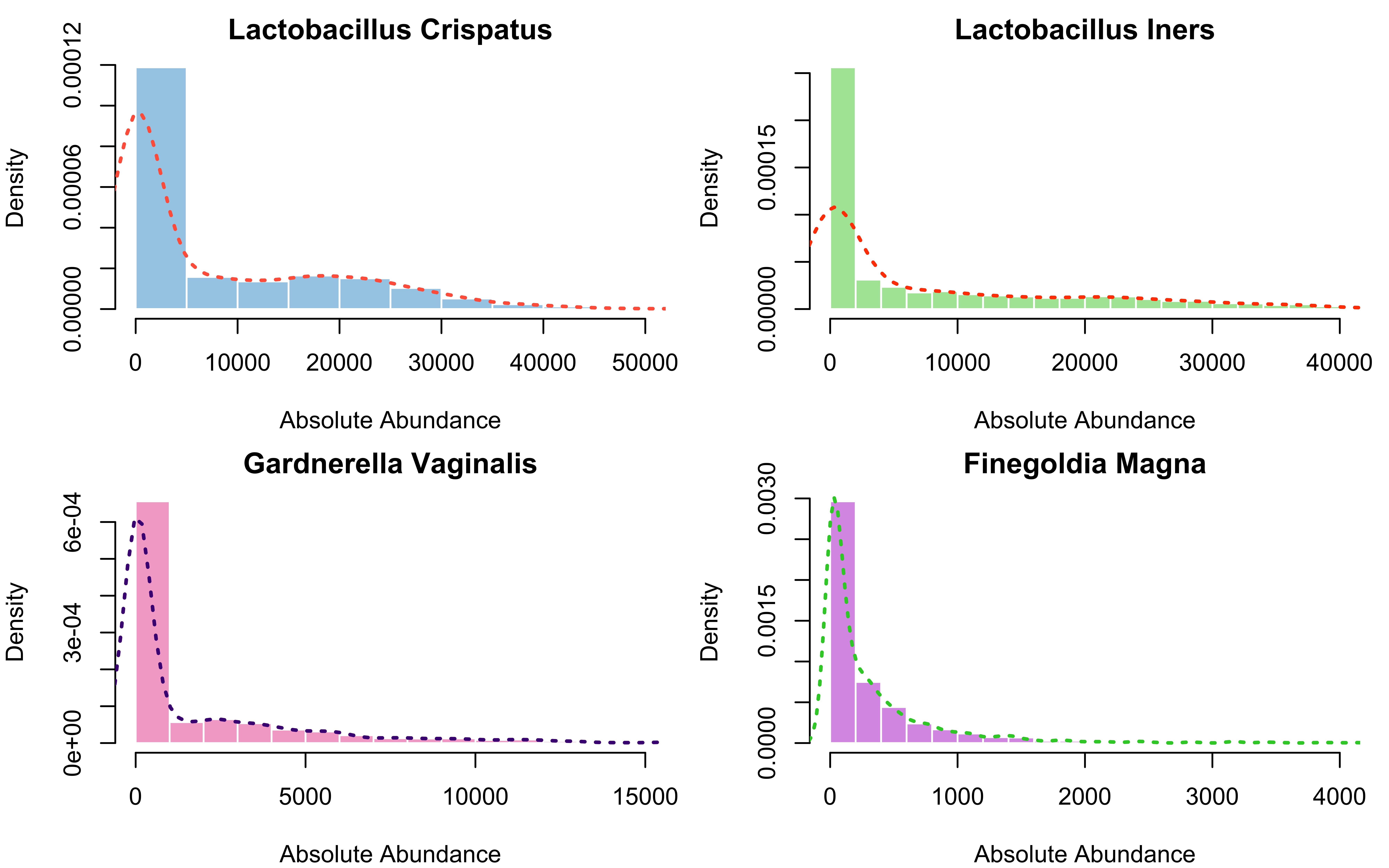}
        \caption{Dispersion and skewness in taxa counts in the CaCHe microbiome dataset.}
        
        \label{fig:disp_plot}   
\end{figure}

The four subplots in  Figure \ref{fig:disp_plot} show how the empirical densities (dotted lines in each figure) of the raw sequence read counts of taxa are highly skewed and vary significantly across taxa. For example, the sequence read counts of \textit{Lactobacillus iners} is highly skewed, whereas \textit{Finegoldia magna} demonstrates less skewness in our data. The horizontal axis of this figure shows how the read counts  of \textit{Lactobacillus crispatus} and \textit{Lactobacillus iners} can be as high as 40,000 to 50,000 per observation, but those for \textit{Gardnerella vaginalis} and  \textit{Finegoldia magna} do not exceed 15,000 and 4,000 respectively.

\begin{comment}

\begin{figure}[H]
    \centering
    \includegraphics[width=0.8\textwidth]{Sequence Depth3.pdf}
    \caption{Differential depths of sequencing}
    \label{fig:diffdepths_plot}
\end{figure}
\begin{figure}[H]
    \centering
    \includegraphics[width=1\textwidth, height = 0.3\textheight]{column_pi.pdf}
    \caption{Proportion of zeros in taxa counts in the study data}

    \label{fig:pi_plot}
\end{figure}

\begin{figure}[H]
    \centering
    \includegraphics[width=1\textwidth, height = 0.3\textheight]{Overdispersion3.pdf}
    \caption{Dispersion in taxa counts in the study data}
    \label{fig:disp_plot}
\end{figure}
\end{comment}

\begin{figure}[H]
    \centering
    \includegraphics[width=1.0\textwidth, height = 0.30\textheight]{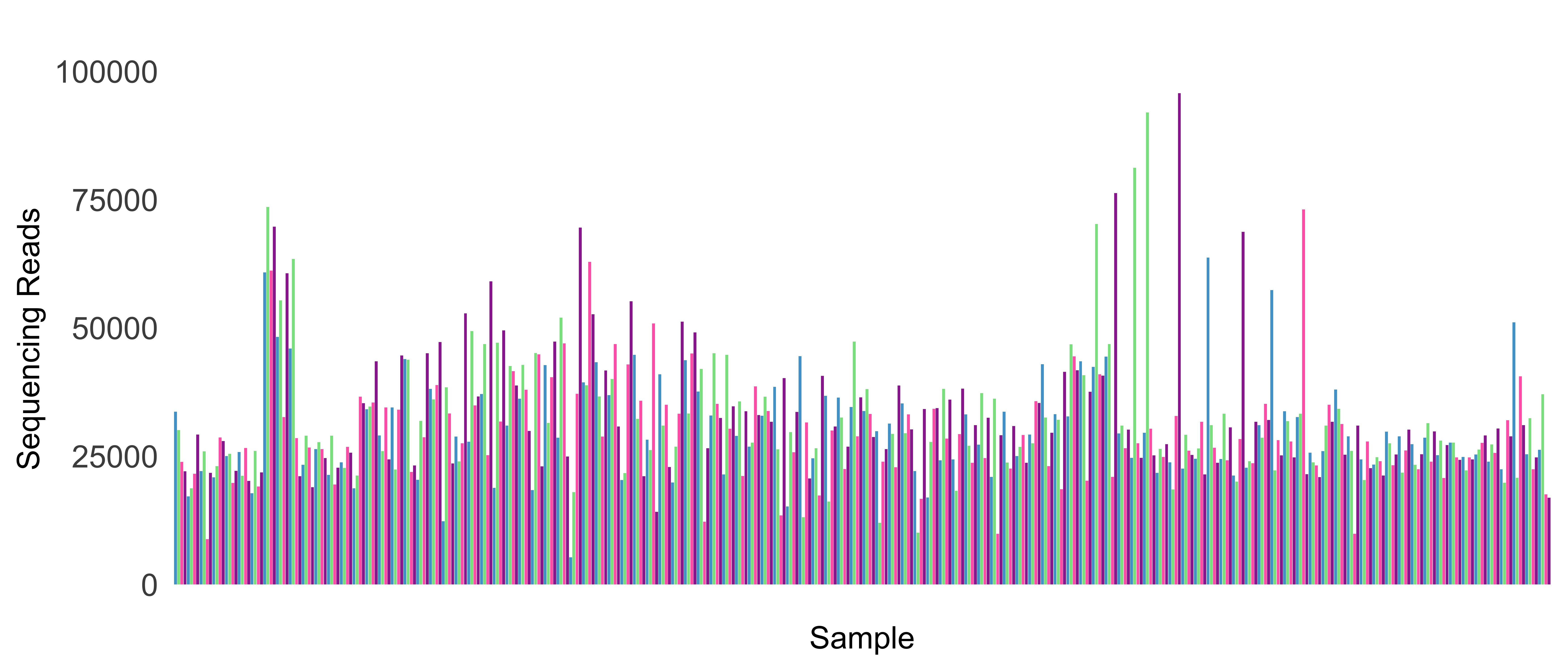}
    \caption{Differential depths of subject-level sequencing in the CaCHe microbiome data.}
    \label{fig:diffdepths_plot}
\end{figure}

Figure \ref{fig:diffdepths_plot} displays the variation in  sequencing depths at the individual subject level, with total sequence counts spanning from  10,000 to almost 100,000 in our dataset. This figure consists of 436 bars where each bar represents the total sequence count of an individual subject.
In the following section, we develop a model for taxa counts addressing all of these three characteristics, namely zero-inflation, overdispersion, and differential depths.  We will call the model a Zero-Inflated Mediator Model.

\section{Zero-Inflated Mediator Model}
\label{sec:mediator-model}

Mediation refers to a process by which the relationship between an independent variable (e.g. exposure) and a dependent variable (e.g. outcome) is influenced by a third variable, known as a mediator \cite{columbia2025causal}. 
Traditional mediation methods (e.g., Baron and Kenny) decompose the total effect into direct and indirect effects based on the linear relationship between the exposure and outcome. These methods rely on assumptions that there are no interactions between the exposure and the mediator, and that no unmeasured confounding factors are present in the analysis \cite{baron_moderatormediator_1986}. Next we clarify the role of confounders after developing a probability model for zero-inflated mediators. 

\begin{figure}[H]
    \hspace{-0cm} % Adjust the value to move further left
    \includegraphics[width=1\textwidth, height=0.30\textheight]{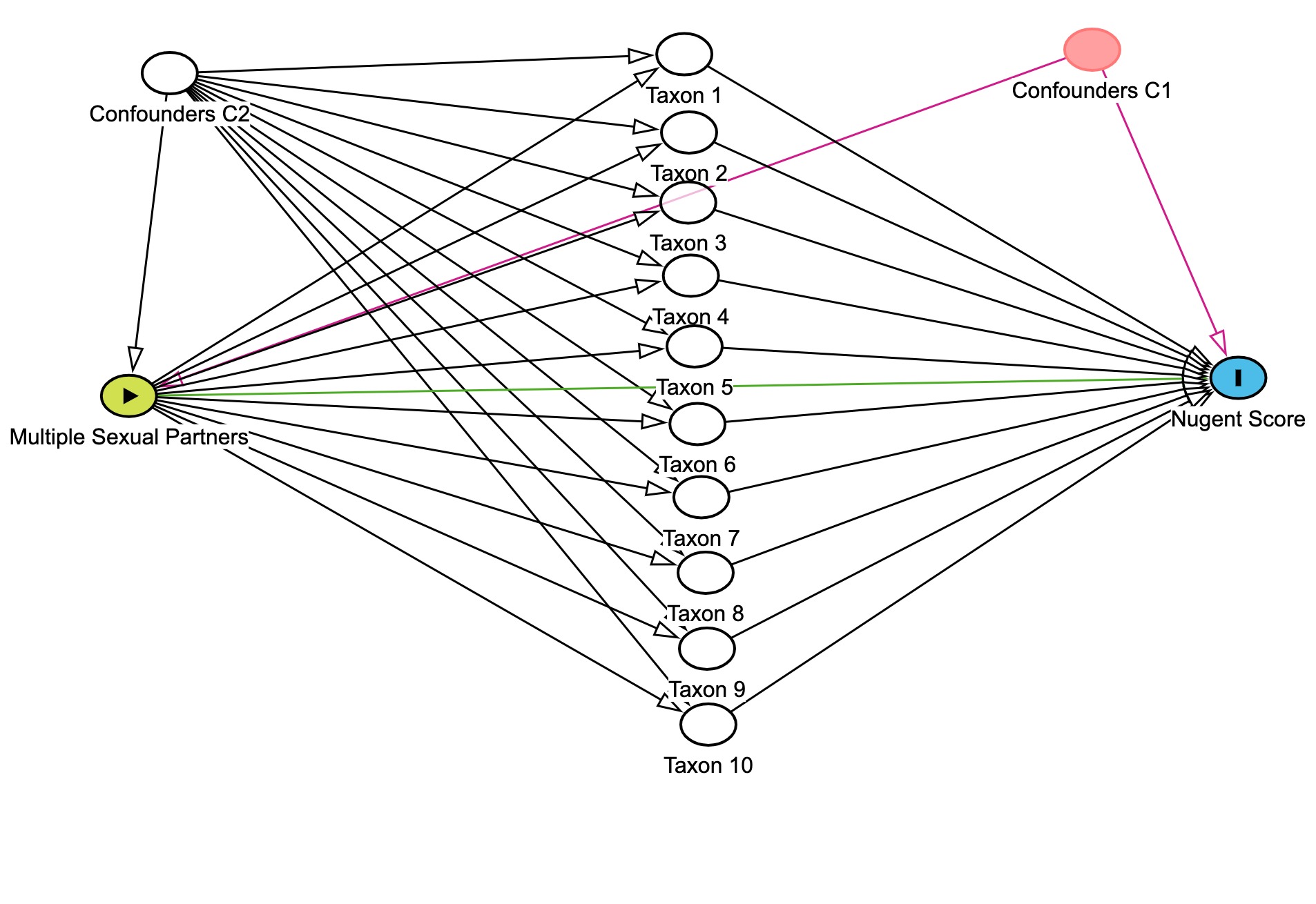}
    \caption{Directed acyclic graph of the mediation model.}
    \label{fig:dag_plot}
\end{figure}

Let $M = (M_1, M_2, ....M_p)$ be a vector of mediators representing multiple taxa in a microbiome dataset. We use the following notations:  $i$ represents a study  subject ($i = 1, \cdots, n$), $j$ represents a mediator ($j = 1, \cdots, p$),  and $M_{ij}$ denotes the raw sequence count of the mediator $j$ for the $i^{th}$ subject. We assume $M_{ij}$ follows a mixed effects zero-inflated generalized linear model (MEZIGLM) with a subject-specific exposure $A_i$, covariate $C_{2i}$ and a random effect $\delta_i$ to account for correlations between varying total sequence reads nested within the subject. We assume that $\delta_i$ follows a normal distribution with mean $zero$ and variance $\sigma_{\delta}^2$. We further assume that pathways do not interact, allowing us to quantify the effect of each pathway in isolation. The conceptual directed acyclic graph (DAG) is described in Figure \ref{fig:dag_plot}. The black arrows from multiple sex partners (MSP) through each taxon to Nugent BV score (i.e. a measure of BV, ranging from 0-10, where higher scores (7-10) are diagnostic of BV) represent the indirect pathways, while the green arrow from MSP to BV represents the direct effect pathway.  C1 and C2 represent confounders affecting the exposure-outcome and exposure-mediator relationships, respectively.

   The MEZIGLM model is expressed below:
\begin{equation}
Pr(M_{ij} = m) =
\begin{cases}
  \pi_j + (1 - \pi_j)\times g(m,\lambda_{ij},\phi_{j}), & \text{if } m = 0, \\
  (1 - \pi_j) \times g(m,\lambda_{ij}, \phi_{j}), & \text{if } m > 0,
\end{cases}
\label{model-mediator}
\end{equation}
where $\pi_j$ represents the likelihood of zero-inflation for the $j^{th}$ taxon which follows a logit  model as given in (\ref{model-mediator-mean}). The function $g$ ($m$, $\lambda_{ij}$,  $\phi_j$), represents a standard distribution for count data (e.g.  Poisson,  Negative Binomial, etc.). 
$\lambda_{ij}$ denotes the conditional mean (given the random effect)   of $g$ which follows a log-linear model  and $\phi_j$, the dispersion parameter follows another  log-linear model given in (\ref{model-mediator-mean}). Note that the dispersion parameter has an inverse relation with the variance of the distribution, meaning a larger dispersion indicates a smaller variance and vice versa \cite{mccullagh1989generalized}.
\begin{equation}
\pi_j = \frac{exp(\beta_{z0j})}{1 + exp(\beta_{z0j})}, \; \lambda_{ij} = exp(\beta_{0j} + \beta_{1j}A_i + \beta_{2j}C_{2i} + log(\zeta_i) + Z_i\delta_i), \mbox{and}\; \phi_j = exp(\beta_{l0j}).
\label{model-mediator-mean}
\end{equation} 
\begin{comment}
    and

\begin{equation}
    \phi_j = exp(\beta_{l0j}).
\label{model-mediator-dispersion}
\end{equation}
\end{comment} 
 The assumption that  $\pi_j$ and $\phi_j$  depends on $j$ is justified as some bacteria tend to be more   abundant or overdispersed in contrast to some of the rarer bacteria with inflated zero counts. $\lambda_{ij}$ is a function of subject-specific exposure $A_i$, covariate $C_{2i}$, an offset parameter $\zeta_{i}$ to adjust for varying sequence reads across subjects.   For convenience, let us define a vector of parameters for the mean component  by  $\bm{\beta}_j$= ($\beta_{0j}$,$\beta_{1j}$,$\beta_{2j})^t$, the zero-inflation parameter by $\beta_{z0j}$, and the dispersion parameter by $\beta_{l0j}$. 
Let $\bm{\beta}$ = $(\bm{\beta}_{z0}, \bm{\beta}_{l0}, \bm{\beta}_0, \bm{\beta}_1,  \bm{\beta}_2)^t, $, where $\bm{\beta}_{z0}=(\beta_{z01}, \dots, \beta_{z0p})^t$, $\bm{\beta}_{l0}=(\beta_{l01}, \dots, \beta_{l0p})^t$, and $\bm{\beta}_{i}=(\beta_{i1}, \dots, \beta_{ip})^t$, $i=1,2,3$, thus, the dimension of $\bm{\beta}$ is $p+p+3p$. Let us denote $\bm{\Theta}= (\bm{\beta}, \sigma^2_{\delta})$, the vector of all parameters with the dimension $5p+1=P$ (say).
 We estimate $\bm{\Theta} $  by the maximum marginal  likelihood method, and denote it by $\hat{\bm{\Theta}}$. The asymptotic  distribution of $\hat{\bm{\Theta}}$ follows a multivariate normal distribution,  i.e. $\hat{\bm{\Theta}}$ $\sim$ $N_P$($\bm{\Theta}$, $\bm{\Sigma}$). This asymptotic distribution will be explored further in  Section \ref{sec:fiducial} to construct fiducial quantities of several pivot statistics.

\noindent {\bf{The Exposure Model}}: Let $A_i$ be a binary exposure, and $C_{1i} = (c_{1i}, c_{2i},\cdots c_{ri})^t$ be a vector of covariates of the $i^{th}$ subject that potentially confound the exposure-outcome relationship. To adjust for this confounding effect, we model the exposure status $A_i$ as a function of covariates $C_{1i}$ as follows: 
\begin{equation}
    logit\; Pr(A = A_i|C = C_{1i}) = \alpha_0 + \alpha_1c_{1i} + \alpha_2c_{2i} + ... \alpha_rc_{ri}.
    \label{exposure}
\end{equation}

Estimates of parameters of Model (\ref{exposure}) will be used to compute a weight function  to adjust for any potential exposure-outcome confounding in the model. 

\noindent {\bf{ The Exposure-Outcome Model}}: Let $Y_i$ be the outcome measure of the $i^{th}$ subject with the exposure $A_{i0}$. We propose the following marginal linear  model to express the relationship between the outcome measure and exposure through  causal pathways.
\begin{equation}
   E(Y_{i M_{1}, \cdots, M_{p}|{\bm{A_i}}})  = \theta+ \theta_0A_{i0} + \theta_1A_{i1} + ... \theta_pA_{ip}.
    \label{outcome-exposure}
\end{equation}
Let ${\bm{A_i}}$= $(A_{i1}, \cdots ,A_{ip})^t$ be an exposure vector of dimension $p$ for the $ith$ subject, where each $A_{ij}$ can take  either a level value $a^*$ (e.g. with no exposure) or $a$ (e.g. at the presence of the exposure). Hence, the elements of the vector $ {\bf{A_i}}$ can be arranged in a total of  $2^p$ different ways. These pseudo vectors of exposures play a crucial role in the estimation process of causal effects.
 Using   Model (\ref{outcome-exposure}), a possible extended dataset is created  for every subject's outcome measure $ Y_i$ with each arrangement of ${\bf{A_i}}$.  Denote a pseudo exposure vector by ${\bf{A_{il}}}$, where $l=1, 2, \cdots, 2^p$, and the corresponding $Y_i$ by $Y_{il}$, even though $Y_{i1} =Y_{i2}, \cdots, = Y_{i2^p}$, however, the suffix $l$ in $Y_{il}$  is important for the right side expression of (\ref {outcome-exposure}). In light of this notation, Model (\ref{outcome-exposure}) can be written as $Y_{i M_{1}, \cdots, M_{p}|{\bm{A_{il}}}}  = \theta+ \theta_0A_{il0} + \theta_1A_{il1} + ... \theta_pA_{ilp} +\xi_{il}$. Parameters in  (\ref{outcome-exposure}) are estimated by the  weighted least square method where weights defined below are functions of the Mediator Model (\ref{model-mediator}) and Exposure Model (\ref{exposure}). Denote the weight function for the $i^{th}$ subject with the $l^{th}$ arrangement by $W_{il}$ as described in \cite{lange_simple_2012, lange2014multiple}, where
 
\begin{equation}
    W_{il} = \frac{P(A = A_{i0})}{P(A = A_{i0}|C = C_{1i})}\prod_{j = 1}^{p}\frac{P(M_{ij} = m| A = A_{ilj}, C = C_{2i}, \delta_i)}{P(M_{ij} = m| A = A_{i0}, C = C_{2i}, \delta_i)}.
    \label{weight-function}
\end{equation}
In (\ref{weight-function}), $A_{i0}$ is the true exposure,  the $i^{th}$ subject was exposed at, and $ A_{ilj}$ is a pseudo exposure for the $l^{th}$ arrangement associated with the $j^{th}$ mediator.  The first fraction in (\ref{weight-function}) addresses exposure-outcome confounding by reweighting observations. This reweighting balances the distribution of confounder variables between the two exposure groups \cite{lange_simple_2012}, similar to how it would be in a randomized trial. The second fraction is calculated using the estimated parameters from the Mediator Model  (\ref{model-mediator}). It distinguishes between the observed mediator value and the value it would have taken in the presence of a counterfactual exposure. If the exposure had no effect on the mediator, the second fraction would effectively be 1. Note that   mediators play an important role in estimating the weight function $W_{il}$ using the corresponding probability function (\ref{model-mediator}). The first part of the numerator, i.e.  $Pr(A=A_{i0})$ is primarily used to stabilize the weight function, and  is obtained by averaging the probability function with respect to $C_{1i}$. 
After estimating the weights, parameters   ${\bm{\theta}} = (\theta, \theta_0, \theta_1, \cdots, \theta_p)'$ in (\ref{outcome-exposure})  are estimated by the  weighted least square technique  denoted by $\hat{\bm{\theta}}$ and will be used  to estimate the causal effect of mediators.

\section{Causal Mediation Analysis}
\label{sec:causal_med}
Causal mediation analysis defines the causal mechanism through which an exposure causes an outcome. It is an extension of the traditional mediation approach which mainly focuses on the association between variables, i.e., how an exposure affects an outcome through a mediator. However, this approach has several limitations such as inaccurate effect decomposition in case of exposure-mediator interaction, inability to handle non-linear models and exposure-mediator or mediator-outcome confounding cases \cite{Schuler2025}.   Addressing these limitations \cite{pearl_3_2010,imai_general_2010,albert_generalized_2011}, the causal mediation method is established under some clear assumptions to identify and estimate the direct and indirect effects.  The approach of analysis can be either parametric or non-parametric, offering flexibility to include both linear and non-linear models. Recent advancements in the causal mediation domain have expanded the scope to incorporate various types of mediators (e.g., binary, continuous) and outcomes (e.g., binary, continuous, or time-to-event) \cite{vanderweele_odds_2010,lange_survival_2011}. Additionally, in the causal mediation framework, numerous methods have been proposed to address challenges such as multiple mediators, exposure-mediator interaction, and mediator-outcome confounding \cite{Valeri2013, vanderweele_mediation_2014}. 

 The standard   approach for causal mediation analysis by  Pearl with two regression models; one to capture the relationship between  exposure and  mediators, and the other one to model the relationship between  exposure and  outcome through the mediators,  encounters challenges like complex integration and conditional effect estimation \cite{pearl_3_2010}. Nested Marginal Structural Model (MSM) by \cite{vanderweele_marginal_2009} enhances robustness against model misspecification, while a method proposed by \cite{van_der_laan_direct_2008} provides a doubly robust estimator for direct effects, and \cite{tchetgen_semiparametric_2012, tchetgen_tchetgen_estimation_2014} improves estimators for indirect effects, facilitating hypothesis testing across various mediators and outcomes.

\subsection{Direct and indirect effects} 

Causal mediation effect defines the causal effect of a treatment on an outcome which is the difference between two counterfactual outcomes.  Consider a binary exposure with symbols $a^*$ and $a$, and let ${\bf{M_{A^*}}}$ be the  value of ${\bm{M}} = (M_1, M_2, \cdots, M_p)^t$ if the exposure vector ${\bm{A_l}}$ is set to  ${\bm{A^*}}=(a^*, a^*, \cdots, a^*)^t$. Let $Y_{a^*, {\bm{M}_{A^*}}}$ denote the  value of $Y$ when the exposure is set at $a^*$ and the value the mediator vector $\bm{M}$ would have taken if $\bm{A}$ were set to  $A^*$.  In a nested counterfactual scenario, $Y_{a^*,M_{\bf{A_l}}}$ denotes the counterfactual outcome  $Y$ if the exposure  is set to $a^*$ and the value the mediator vector $\bm{M}$ would have taken if $\bm{A}$ were set to ${\bf{A}_l}$. 

The total effect (TE) of the exposure on the outcome can be decomposed into two parts,   natural direct effect (NDE),  one that directly goes from the exposure  to the outcome $Y$, and  natural indirect effect (NIE), that goes from the exposure  to the outcome $Y$ indirectly by passing through the mediator vector $\bm{M}$. In a given experimental setup, NDE is defined  as the expected difference between two counterfactual outcomes measured at two different exposure levels (i.e $a$ and $a^*$) but mediator vectors are measured at the same specific arrangement (say ${\bm{M_{A^*}}}$). Likewise, NIE is defined as the expected difference between two  counterfactual outcomes measured at the same exposure level but mediator vectors are measured at two specific arrangements of $\bm{A}$, say ${\bm{{\bm{A_{h}^*}}}}$ and ${\bm{A^*}}$.
\begin{equation}
    NDE  = E_{\delta} [E(Y_{a, {\bm{M_{A^*}}}} -  Y_{a^*, {\bm{M_{A^*}}}}|\delta_i)], \; NIE  = E_{\delta}[E(Y_{a^*, {\bm{M}}_{{\bm{A}}_h^*}}  - Y_{a^*, {\bm{M_{A^*}}}} | \delta_i)],
    \label{nie}
\end{equation}

 where $\bm{A}_h^*$ = $(a^*,  \cdots, a^*, a, a^*, \cdots, a^*)^t$, i.e. the $h^{th}$ element of $\bm{A}_h^*$ is $a$, and all other elements are $a^*$. The term "natural" signifies the value that the mediator vector $\bm{M}$  would naturally assume if the exposure $A$ were set to $a^*$. The assumptions needed to estimate direct and indirect effects are: (i) there are no unmeasured confounders affecting the exposure-outcome (A-Y) relation, (ii) there are no unmeasured confounders affecting the mediator-outcome relation ($\bm{M}$ -Y), (iii) there are no unmeasured confounders affecting the exposure-mediator relation (A-$\bm{M}$), and (iv) there are no mediator-outcome (M-Y) confounders that are affected by the exposure $A$ \cite{lange2014multiple, vansteelandt_imputation_2012}. Further discussion on the validity of assumptions (i) to (iv) in relation to our example data will be presented in Section 7.
\section{ The Fiducial Approach}
\label{sec:fiducial}
The fiducial approach, introduced by Ronald A. Fisher, is a statistical framework for deriving inferential quantities based on the observed data and sampling distributions,  without relying on prior distributions. There are several ways to construct a fiducial quantity or a generalized pivotal quantity for a parameter, which can subsequently be used for hypothesis testing or constructing confidence intervals \cite{Weerahandi1993, Cisewski2012, Hannig2016, Mathew2013}. We state that a fiducial quantity, say  $\widetilde{\bm{\gamma}}$ of a parameter ${\bm{\gamma}}$, is a function of the observed data, and a pivot statistic should maintain two required conditions: (i) given the observed data, the distribution of $\widetilde{\bm{\gamma}}$ is free of any unknown parameters, and (ii) if random quantities in the pivot statistic are replaced by the corresponding observed quantities, $\widetilde{\bm{\gamma}}$ simplifies to $\bm{\gamma}$ \cite{bhaumik2021generalized}.

Utilizing the sample data and exploring distributions of estimated model parameters, our proposed approach generates weights defined in (\ref{weight-function}) multiple times. Using the generated weights, the approach  develops an empirical density,  the basis for constructing a confidence interval, known as the generalized confidence interval (GCI) for the NIE. The GCI is derived following the highest empirical density concept \cite{Kruschke2014}.  To   construct  fiducial quantities for  parameters of the MEZIGLM model,  we are going to use an approach given in \cite{krishnamoorthy2006}. We clarify two notations in general:   $\hat{\gamma}$ is used when a parameter $\gamma$ is estimated from the given data, and $\hat{\hat{\gamma}}$ is used for a random observation when it  is drawn from the asymptotic distribution of $\hat{\gamma}$. Thus, $\hat{\gamma}$ is a realization of $\hat{\hat{\gamma}}$. In what follows, we will  repeatedly use a result  that if $F(\bm{\gamma})$ is a real-valued function of $\bm{\gamma}$, and $\bm{\tilde{\gamma}}$ is a fiducial quantity of $\bm{\gamma}$ then $F(\bm{\tilde{\gamma}})$ is a fiducial quantity of $F(\bm{\gamma})$. Next we proceed towards the construction of a fiducial quantity of NIE. 
 
Let $\bm{\Theta}$ be the vector of all parameters, i.e. $\bm{\Theta}$ = $(\bm{\beta} \; \sigma_{\delta})^t$ is of dimension $P$ as mentioned before. Assume that   $\hat{\bm{\Theta}}$ is a consistent estimate of $\bm{\Theta}$, and $Cov(\hat{\bm{\Theta}})$ = $\bm{\Sigma}$.  Let $\bm{S}^*$ be the inverse of the sample Fisher information matrix which is a consistent estimator of $\bm{\Sigma}$.  However, the distributional property of such an estimate  is not well developed. Using the parametric bootstrapping approach, we estimate $\bm{\Theta}$  a total of $N+1$ times, such estimates, say ${\bm{\hat{\Theta}}_1}$, $\bm{\hat{\Theta}_2}$, $\cdots,$ $\bm{\hat{\Theta}_{N+1}}$ are independent. Let $\bm{S}_N = \bm{S} = \frac{\sum_{l=1}({\bm{\hat{\Theta}}_l} - {\bm{\bar{\hat{\Theta}}}})({\bm{\hat{\Theta}}_l} - {\bm{\bar{\hat{\Theta}}}})'}{N} $, where ${\bm{\bar{\hat{\Theta}}}} = \frac{\sum_{l} {\bm{\hat{\Theta}}_l}}{N+1}$.  Both $\bm{S}^*$ and $\bm{S}$ are consistent estimates of $\bm{\Sigma}$. $N$ is estimated using the equivalence of matrix norms (e.g. $L_1$ or $L_2$ norms on the difference of corresponding eigen values of 
 these two matrices). Note that $N\bm{S}$ $\sim$ $W_P (N, \bm{\Sigma}$), a Wishart distribution of dimension $P$, degrees of freedom $N$, and parameter matrix $\bm{\Sigma}$, thus $E(\bm{S})= \bm{\Sigma}$. Note that $|\bm{S}^* - \bm{S}_N|$ $\xrightarrow{P} $ $\bm{0}$, in probability. So, we can assume $\bm{S}^*$ is a realization of $\bm{S}$ for a suitably chosen $N$. Let us term $N$ as the equivalence number, we will discuss in details how to determine the value of $N$ in the Simulation Study Section \ref{sec:simulation}.
 
 Let $\bm{T}$ be the Cholesky decomposition of $N\bm{S}$, that is $N\bm{S}$ = $\bf{TT'}$, where $\bf{T}$  is a lower triangular matrix with positive diagonal elements. Similarly, let $\bm{\Gamma}$ be the  Cholesky decomposition of $N\bm{\Sigma}$, and  $N\bm{\Sigma}$ = $\bm{\Gamma \Gamma'}$. Let $\bm{U}$ = $\bm{\Gamma^{-1}}$ $\bf{T}$ = $((u_{ij}))$, a lower triangular matrix. The distribution of $\bm{UU'}$ = $\bm{\Gamma}^{-1} \bf{T} \bf{T'} \bm{\Gamma'}^{-1}$ is Wishart (\cite{Muirhead1982}), i.e.

\begin{equation}  
  \bm{UU'} \sim W_P(N, I_P),  \mbox{ and all elements of $\bm{U}$  are independent,}
  \label{wishart}
\end{equation} 
where $u_{ii}$ $\sim$ $\chi^2_{N-i}$, $i=1,2, \cdots, P$, and $u_{ij}$ $\sim$ $N(0, 1)$, $i > j$.
 \begin{itemize}
 \item{Fiducial quantity for $\bm{\Sigma}$}:
 As mentioned before, $\hat{\Theta}$ $\xrightarrow{P}$  $\Theta$, and $\bm{S}$ $\xrightarrow{P}$  ${\Sigma}$. Further, let $\bf{t}$ be the Cholesky decomposition of $N\bm{S^*}$, then $\bf{t}$ is a sample realization of $\bf{T}$. We do not assume any distribution on $\bm{S^*}$.
  Denote $\bf{B}$ = $((b_{ij}))$ = $\bf{t}$ ${\bm{U}}^{-1}$, thus, it  is a lower triangular matrix with all  elements $b_{ij}=0$, for $i<j$
 and is a fiducial quantity for $\bm{\Gamma}$.
 %\begin{equation}
 %\bf{B} = \bf{t}{\bm{U}}^{-1} \; \;  \mbox{is a fiducial quantity for $\bm{\Gamma}$}.
 %\label{fiducial for Sigma}
 %\end{equation}
 As $\bf{t}$ is a statistic based on the  observed data, and the distribution of $\bm{U}$ does not depend on any parameter, therefore (i) given $\bf{t}$, the distribution of $\bf{B}$ is free of any unknown parameters. Next, (ii) $\bf{B}$ =  $\bf{t}$${\bm{U}}^{-1}$ = $\bf{t}$$\bf{T^{-1}}$$\bm{\Gamma}$ = $\bm{\Gamma}$,  when the observed quantity $\bf{t}$ and the random draw $\bm{T}$ are equal. Thus, $\bf{B}$ satisfies both conditions to be a fiducial quantity of $\bm{\Gamma}$. 
 ${\bf{BB'}}/N$ is a fiducial quantity of $\bf{\Sigma}$, we denote it  by $\widetilde{\bf{\Sigma}}$.  
 
\item{Fiducial quantity for $\bm{\Theta}$}
\begin{equation}
    \Tilde{{\bm{\Theta}}} = \hat{\bm{\Theta}} - {\bm{B}^{-1/2}} {\bm{\Gamma}}^{1/2}
    (\bm{\hat{\hat\Theta}} -\bm{\Theta}) = \hat{\bm{\Theta}} - {\bm{B}^{-1/2}} \bm{Z},
\label{model-betaj-fiducial}
\end{equation}
where $ \bm{Z}$ follows a multivariate normal distribution.  We see in the last expression of  (\ref{model-betaj-fiducial}) that (i) given data (i.e. $\hat{\bm{\Theta}}$ and $\bm{t}$), the distribution of $\Tilde{{\bm{\Theta}}} $ does not depend any unknown parameter, 
 (ii)  when the random estimates  become equal to the observed estimates then ${\bm{B}^{-1/2}}$ becomes $ {\bm{\Gamma}}^{1/2} $, and  $\hat{\bm{\Theta}}$ becomes  $\bm{\hat{\hat\Theta}} $, as a results  $\Tilde{{\bm{\Theta}}}$ becomes ${\bm{\Theta}}$. Thus, $\Tilde{{\bm{\Theta}}}$ is a fiducial quantity for ${\bm{\Theta}}$.
Drawing random samples $\bm{Z}$ from a normal distribution $N(\bf{0},\; \bf{I}_P)$, we can generate the fiducial quantity $\Tilde{{\bm{\Theta}}}$ as many times as we want. Decomposing ${\bm{\Theta}}$, we get fiducial quantities for both $\bm{\beta}$ and $\sigma_{\delta}$. For  negative estimates of $\sigma_{\delta}$,  we square  to get a positive estimate of $\sigma^2_{\delta}$.  
\item{Fiducial quantities for $\pi_j$ and $\phi_j$}: 
using $\Tilde{{\bm{\Theta}}}$, we construct fiducial quantities for $\pi_j$ and $\phi_j$ as follows:  
 \begin{equation}
 \Tilde{\pi_j} \sim \frac{exp(\Tilde{\beta}_{z0j})}{1 + exp(\Tilde{\beta}_{z0j})}\; \;  \mbox{and}\; \; \Tilde{\phi_j} \sim \frac{exp(\Tilde{\beta}_{l0j})}{1 + exp(\Tilde{\beta}_{l0j})}.
 \label{fiducial_pi-phi}
\end{equation}
\end{itemize}

\subsection{Statistical inferences  for  mediators} 
 Computation of the standard error of the estimated NIE is challenging, as weights in (\ref{weight-function}) are computed only once using a given dataset, thus the process ignores the standard error of the estimated weights.  Sobel noted that results for mediation effects in smaller samples are underdeveloped, and the standard delta method  can yield biased estimates \cite{stone_robustness_1990}. Non-parametric bootstrapping (NPB) addressed several limitations of the delta method without providing any specific sampling distribution; however, that can be computationally intensive and may suffer from convergence issues for small samples \cite{briggs2007indirect}.  Existing methods for construction of confidence interval for the  NIE using zero-inflated data are conservative in general, and NPB fails to produce satisfactory results in terms of coverage probability, particularly for small samples. In order to address these concerns, we investigate  the fiducial approach in the following section.
  
We demonstrate how to utilize fiducial estimates of parameters to derive a set of weights corresponding to each dataset before estimating the direct and indirect effects. Let us generate K fiducial samples, where the $k^{th}$ sample provides  $\Tilde{\pi}_{jk}$, ${\Tilde{\phi}}_{jk}$, ${\Tilde{\beta}_{0jk}}$, ${\Tilde{\beta}_{1jk}}$ and ${\Tilde{\beta}_{2jk}}$. The empirical Bayes estimate of $\delta_i$ and the fiducial quantities of parameters of the $k^{th}$ sample are now used to compute $ Pr(M_{ij} = m)$ in (\ref{model-mediator}), and denote it by $Pr(M_{ijk}) $. Next, for each $k$, we compute the  weight function (denoted by $W_{ilk}$) in (\ref{weight-function}).  Denoting $E(Y_{i|{\bm{A_i}}})  = \theta+ \theta_0A_{i0} + \theta_1A_{i1} + ... \theta_pA_{ip}$, and the  outcome measure of the $i^{th}$ subject by $Y_i$, we estimate  parameters $(\theta, \theta_0, \theta_1, \cdots, \theta_p)$ by the weighted least square (WLS) method with weights $W_{ilk}$ as follows:

\begin{equation}
    \text{argmin}_{\theta} \sum_{i=1}^{n} \sum_{l=1}^{2^p}{W_{ilk}} (Y_i - Y_{i|{\bm{A_{il}}}})^2.
\label{model-weighted-outcome}
 \end{equation}
We derive a fiducial quantity of the estimated set of parameters of the $k^{th}$ sample denoted by  $(\hat{\theta_k}, \hat{\theta}_{0k}, \hat{\theta}_{1k}, \cdots, \hat{\theta}_{pk}/ {\hat{\delta}_i})$ using a similar approach as illustrated in \ref{model-betaj-fiducial}. The fiducial quantity is then used to compute the conditional NDE and NIE of the $j^{th}$ mediator (i.e. microbiome) following expressions given  (\ref{nie}). For each generated sample, we estimate both NDE and NIE; thus, we get a total of $K$ estimates for each of those.  Denote the fiducial empirical density of the NIE of the  $j^{th}$ mediator by $d_{\mathcal{NIE}_j}(.)$ which is estimated from the $K$ generated fiducial estimates. In the following, we use the estimated 95\% fiducial highest density interval and call it as the {\it{generalized confidence interval}} (GCI),  and the fiducial mode as the point estimate of  $\mathcal{NIE}_j$.

\noindent\textbf{100$(1-\alpha)$\% GCI  by the Fiducial Highest Density (FHD):} the 100$(1-\alpha)$\%  FHD interval for $\mathcal{NIE}_j$ is an interval $[L_j, U_j]$ such that: $\int \limits_{L_j}^{U_j} d_{\mathcal{NIE}_j}(x) \delta x = 1-\alpha$, and minimizes the length of interval, i.e. minimizes $\text{argmin}_{(U_j, \; L_j)}$  $|U_j - L_j|$.  The endpoints $L_j$ and $U_j$ are determined in such a way that it ensures  the interval captures the highest density region.
\textbf{Fiducial Mode:} the fiducial mode $\widehat{\mathcal{NIE}_j}$ is the value of $\mathcal{NIE}_j$ that maximizes the fiducial density function; i.e.,  $\widehat{\mathcal{NIE}_j} = \arg\max_{NIE} d_{\mathcal{NIE}_j}(x)$. This point estimate $\widehat{\mathcal{NIE}_j}$ represents the value of $\mathcal{NIE}_j$ with the highest probability density under the fiducial distribution. To make the   {\it{generalized confidence interval}} unconditional, we numerically integrate the FHD using the distribution of $\hat{\delta_i}$ with the estimated variance $\hat{\sigma}^2_{\delta}$. As we see, the fiducial approach uses the empirical density of the fiducial pivotal quantity, contrary to its competitors (e.g. delta method) that often rely on the normal density. For a skewed distribution of the fiducial quantity, certainly our proposed approach has an edge over, which we will see in the Simulation Section \ref{sec:simulation}. 

\noindent\textbf{Generalized p-value:} The fiducial empirical density further can be used to test $H_0:\; NIE= \kappa$, computing the generalized $p$ value using  the null value of $\kappa$.  We will use the generalized p-value approach for testing a particular NIE in the Data Analysis Section \ref{sec:data_analysis}.

\subsection{Algorithm}
\label{algorithm1}
% the next one is to generate 

\begin{algorithm}[h]
\caption{Algorithm for generating Fiducial Confidence Intervals}
\begin{algorithmic}
\STATE{\textbf{Mediator Model:}}
\STATE{{\textbullet} Use Equation \ref{model-mediator}} to fit the mediator model and estimate $\pi_j$, $\phi_j$,$\beta_j$, $\sigma_{\delta}^2$, and $\Sigma$.
\STATE {\textbf{Generate K fiducial quantities for the MEZIGLM mediator model parameters:}} 
\STATE {{\textbullet} For $\boldsymbol{\Sigma}$: Follow the procedure outlined in Section \ref{sec:fiducial} to generate fiducial quantity for covariance matrix $\Sigma$.}
\STATE {{\textbullet} For $\boldsymbol{\beta_j}$ and $\boldsymbol{\sigma_{\delta}^2}$: Use Equation \ref{model-betaj-fiducial} to generate fiducial quantity for $\beta_j$} and $\sigma_{\delta}^2$.
\STATE {{\textbullet} For $\boldsymbol{\pi_j}$ and $\boldsymbol{\phi_j}$: Use Equation \ref{fiducial_pi-phi} to generate fiducial quantity for $\pi_j$ and $\phi_j$}.

\STATE {\textbf{Computation of fiducial weights:}}
\STATE {{\textbullet} \textbf{Replicate dataset}: Replicate dataset as described in \cite{lange_simple_2012} to create auxiliary variables representing counterfactual exposures in the model}.
\STATE {{\textbullet} \textbf{Compute weights for fiducial parameters:} Calculate weights for each subject corresponding to each set of fiducial parameters $\tilde{\pi_j}$, $\tilde{\phi_j}$, and $\tilde{\beta_j}$} using Equation \ref{weight-function}.
\STATE {{\textbullet} Repeat the following steps for each set of fiducial weights}.
\STATE {\textbf{Outcome Model:}} 

\setlength{\baselineskip}{0pt}
\begin{itemize}
\setlength{\itemsep}{0pt}
    \item Estimate the natural direct and indirect mediation effect by solving the weighted equation \ref{model-weighted-outcome}.
    \item Generate fiducial quantities for each mediation effect using the same concept as illustrated in Equation \ref{model-betaj-fiducial}. 
\end{itemize}

 \STATE \textbf{Generalized confidence interval estimation:} 
 \setlength{\baselineskip}{0pt}
\begin{itemize}
\setlength{\itemsep}{0pt}
    \item Compute the GCI corresponding to each indirect effect, as illustrated in Section 5.1.
\end{itemize} 

\end{algorithmic}
\end{algorithm}

\section{Simulation Study}
\label{sec:simulation}

To evaluate the performance of our approach, we simulated data under various conditions, including different levels of zero inflation, dispersion, and number of mediators, effectively addressing the complexities present in microbiome data. We used a MEZINBM, suitable for the microbiome data in mediation analysis. Simulated results of the fiducial, standard delta, and NPB approaches were compared for different levels of zero inflation and presented in Figure \ref{fig:zi_comparison}. In addition, simulations were conducted across a wide range of dispersion parameters. For all scenarios, the fiducial approach repeatedly outperformed the other two approaches both in terms of sensitivity and coverage probability. Additionally, we show in Figure \ref{fig:mult_med_comparison} that the fiducial approach maintains the target coverage and true detection rate for multiple mediators in medium to large sample size cases.

\subsection{Designs for simulations}

Parameters used in this simulation study were meticulously chosen to reflect the CaCHe microbiome data.  For example, observing the zero inflation rate from 0.10 to 0.90 in Figure (\ref{fig:pi_plot}), we considered three distinct values of $\pi = 0.2, 0.4,$ and $0.6$ for our simulation setup. Similarly, the median dispersion parameter for taxa in the CaCHe data was $0.458$, hence, we considered three distinct values of $\phi=0.5, 1.0$, and $ 10.0$ for this simulation. We set  $\beta_0=-3.0, \beta_1= 0.6, \beta_2= 0.5$,  the parameters  described in (\ref{model-mediator-mean}).  These values were obtained by adjusting for an offset of \textit{log(total sequence count)} for each   subject in the mediator model. In addition, we took $Z_i=1$, and assumed $C_{1i} \sim N(0, 1)$, $C_{2i} \sim N(0, 1)$ , $C_{3i} \sim N(0, 1)$ and $\delta_i \sim N(0, 0.1)$. The standard normality assumption on $C_{ui}$, $u=1, 2, 3$ is based on the idea that covariates can be standardized once their values are obtained. The binary exposure $A_i$ was generated using a logarithmic odds model as a linear function of a continuous covariate $C_{2i}$ with coefficients ($\alpha_0$, $\alpha_1$) = $(0.25, -0.50)$ as denoted in Model [\ref{exposure}].
We assume that  subject specific continuous outcome $Y_i$ follows the model,  $Y_i = \gamma_0 + \gamma_1A_i + \sum_{j=1}^p\gamma_{2j}M_{ij} + \gamma_3C_{3i} + \epsilon_i,\; \mbox{p=1, 3, 5}$.      $Y_i$ was generated as a function conditioning on the exposure $A_i$, mediator vector  $(M_{i1}, ...\; M_{ip})$ and additional covariates $C_{3i}$ with coefficients ($\gamma_0$,$\gamma_1$, $\gamma_3$) = (1.5, 2.0,  1.5) respectively, and  $\gamma_{2j}$ was generated from $N(0.9, .01)$. Searching the literature \cite{abou_chacra2024,subramaniam2016, vodstrcil2015}, we found multiple microbiome studies were performed with sample sizes in the range of 40 to 300, hence for this simulation,  we considered sample sizes $n = 20,40,80,200, 300$ and three sets of mediators $j = 1,3$, and $5$. For each parametric combination mentioned above, the equivalence number $N$  was determined and found to be in between $600$ and $950$. For example, for the median dispersion parameter of $\phi = 0.458$, and the zero inflation parameter $\pi=0.50$ , the equivalence number $N$ is $616$  with the corresponding $L_2$ distance  $0.00180$. 

Each procedure was repeated 10,000 times to construct the confidence interval. All simulations were conducted assuming a single binary exposure $A_i$ for \textit{sp6mosm}, and a continuous outcome $Y_i$ for the \textit{NugentBV} score.  The mean of the MEZINBM $M_{ij}$ was generated for  the given exposure $A_i$ and covariate $C_{2i}$  using the model specified in [\ref{model-mediator-mean}].  Exposure ($A_i$), mediator ($M_{ij}$), and outcome ($Y_i$) were generated using standard R functions such as \textit{rnorm}, \textit{rnbinom}, and \textit{rchisq}. The \textit{GLMMTMB} package was used  to estimate parameters of the MEZINBM model.   Simulations were executed in parallel on a multi-core setup with 200GB of RAM, utilizing the \textit{doParallel} package in R. The computational time for the NPB  was significantly more than that of the fiducial or  delta method. 

 \noindent For each approach, i.e. NPB, delta, and fiducial, we evaluated the coverage probability and width of the corresponding confidence interval for the direct and indirect effects. In addition, we calculated sensitivity, the proportion of times the method was able to capture a significant positive or negative effect (i.e. the confidence interval does not include zero) based on 10,000 repetitions. Sensitivity is a crucial metric for applied scientists to assess the significance of specific biological pathways in mediation analysis. 
 
 Using the MEZINBM model and the aforementioned assumptions of random quantities, the following expressions for the NDE and NIE were derived: $NDE=\gamma_1$ and $NIE = \gamma_{2j}(1-\pi_j)\left[\exp(\beta_{0j} + \beta_{1j} + 0.5\beta_{2j}^2\sigma_{C_2}^2 + 0.5\sigma_{\delta}^2) - \exp(\beta_{0j}  + 0.5\beta_{2j}^2\sigma_{C_2}^2 + 0.5\sigma_{\delta}^2) \right]$. The value of an NIE calculated using this expression is referred to as the gold standard value (GSV). The taxon-specific GSV of the NIE plays a crucial role in determining whether the GCI (or CI for NPB) includes the NIE. Additionally, it aids in calculating the bias of the NIE. Throughout our simulations, we observed both negative and positive biases; however, these biases were very close to zero.
 
 As illustrated in Figure \ref{fig:zi_comparison}, we found that  coverage probabilities (green lines in the top panel) of confidence intervals for the indirect effect determined by the fiducial approach were always around 0.95   close to the target dashed line for n=80, 200, and 300.  The performance of the delta method (blue lines in the top panel) was comparatively weaker, never exceeded 0.95 for any combination of sample size or zero inflation parameter. The coverage probabilities of NPB (orange lines in the top panel) ranged from  0.90 to 0.95 for $n=200$ and $300$. For smaller samples (i.e. $n \leq 80$) both NPB and the delta method displayed poor performance in terms of  coverage probability. Corresponding confidence widths (CWs) are presented in the bottom panel. 
The CWs of NPB (orange bars) were consistently wider compared to those of the fiducial and  delta methods. The delta method had shorter CWs, which justified its smaller coverage probabilities.  The fiducial approach, positioned in the middle (green bars),  achieved the target coverage probability of 0.95 for $n=200$ and $300$. The take-home message is that both NPB and fiducial perform well for large samples, but fiducial has an edge over the NPB for small samples, and the delta method fails to produce desirable results.

\noindent {\bf{Model misspecifications:}} To examine the impact of misspecification of models,  we analyzed the mediator data using a mixed effects zero-inflated Poisson model (MEZIPM) even though data was generated by the MEZINBM as discussed before. The coverage probability was much less than what we expected for a mediator; for example,  it was 0.87 for $\pi=.20$, 0.86 for $\pi=0.40$, and 0.84 for $\pi = 0.60$, for a fixed sample size  $n=200$. 

\subsubsection{ Simulated coverage probabilities and  widths of  confidence intervals}

\begin{figure}[H]
  \includegraphics[width = 1\textwidth, height = 0.5\textheight]{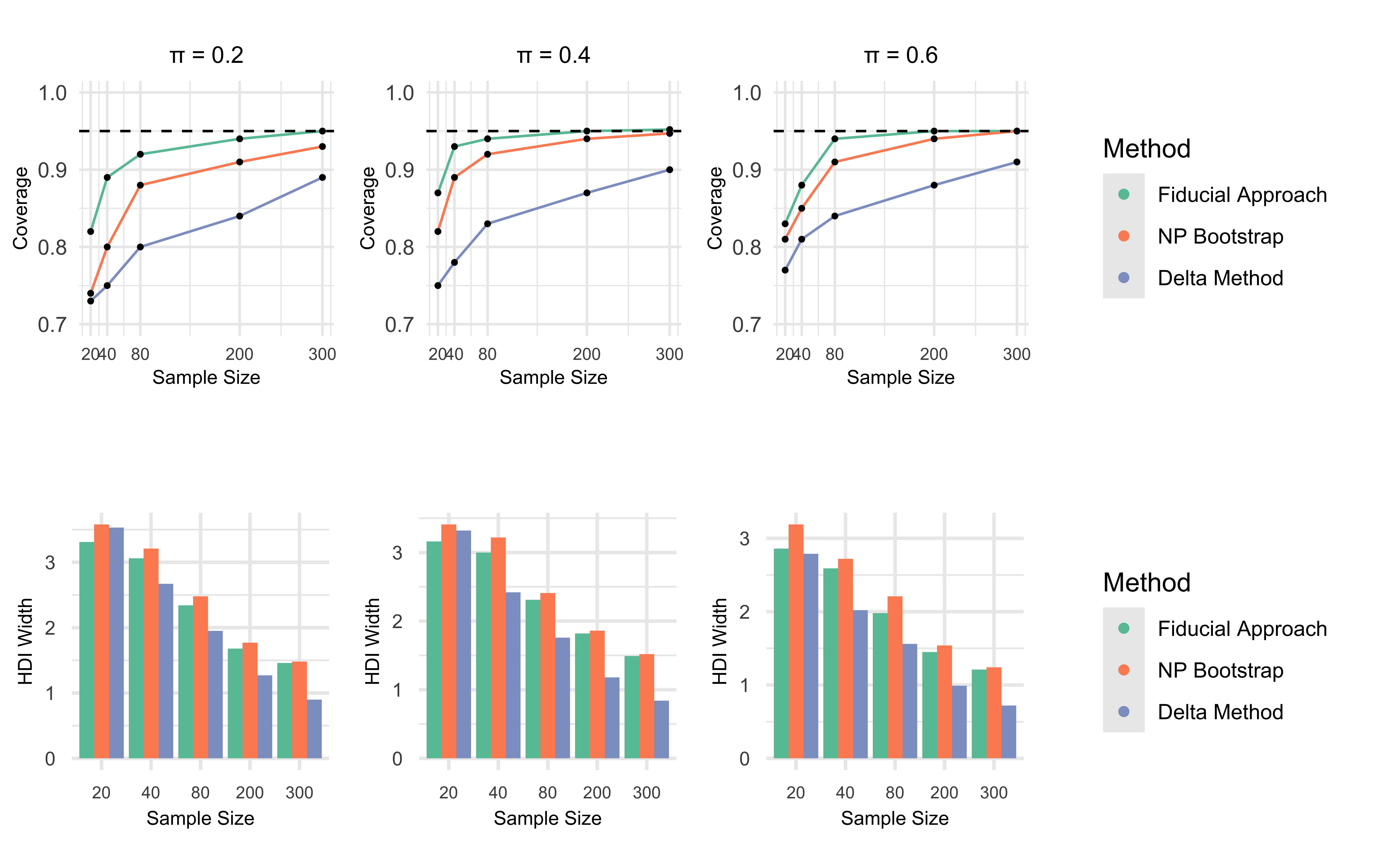}
    \caption{Comparing coverage probabilities and widths of confidence intervals  for the natural indirect effect measured by the fiducial approach, NPB and delta method for a single mediator with different zero inflation levels with a fixed dispersion parameter of 1.0. The black dashed line represents the 95$\%$ coverage line.}
    \label{fig:zi_comparison}
\end{figure}

Similar results were obtained for three values  0.5, 1.0,  10.0 of the dispersion parameter with a fixed zero inflation level of 0.2. The fiducial approach performed the best compared to the other two approaches for small samples; however, for large samples, both the fiducial and NPB approaches performed well, while the delta method remained the weakest performer across all combinations of parameters.

\begin{figure}[H]
    \centering
    \includegraphics[width = 1.1\textwidth, height = 0.40\textheight]{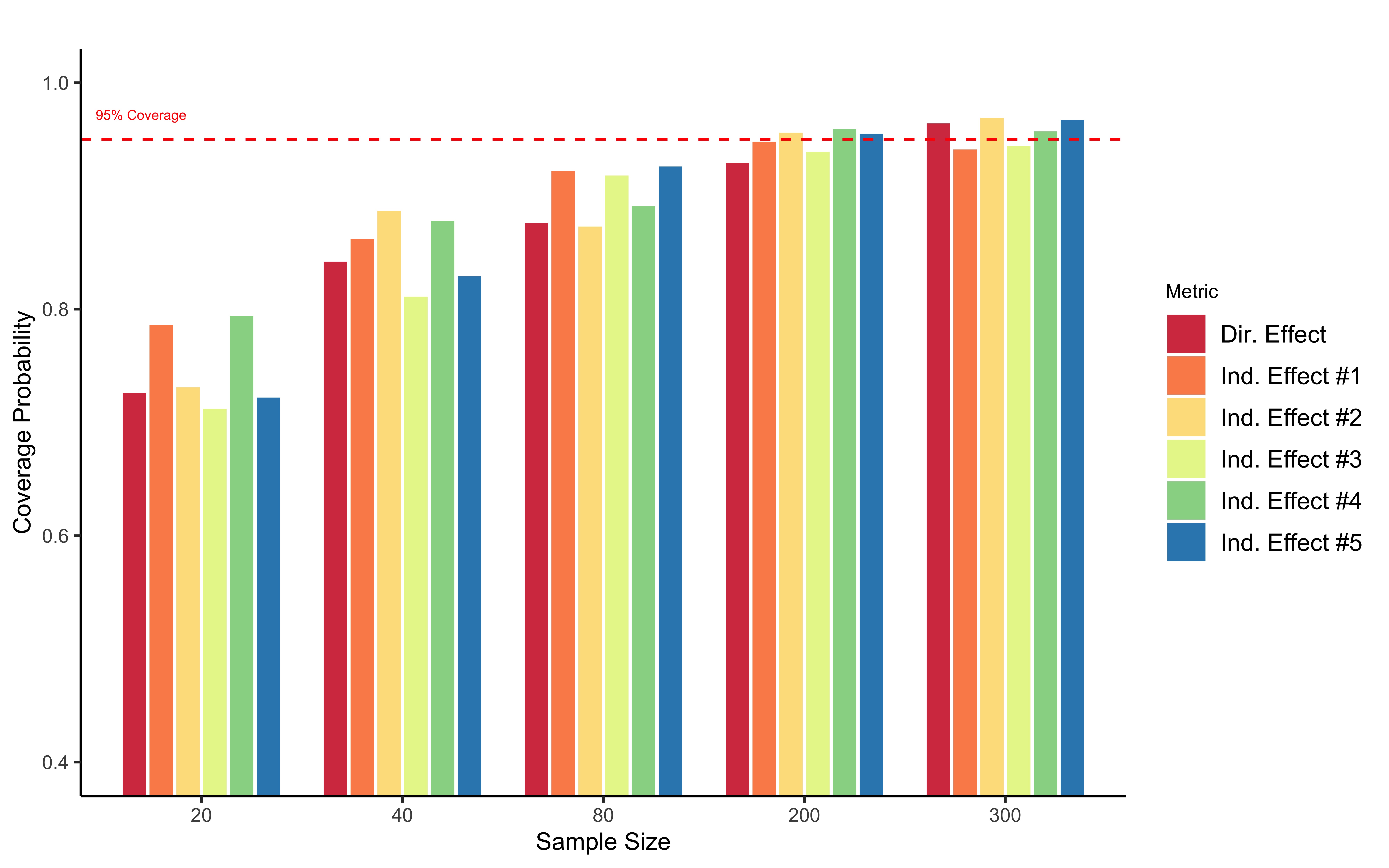}
    \caption{Coverage probabilities of direct and indirect effects by the fiducial approach for multiple mediators incorporating between taxa correlations with varying sample sizes. The red dashed line represents the 95$\%$ coverage line.}
    
    \label{fig:mult_med_comparison}
\end{figure}

 Coverage probabilities of both direct and indirect effects of multiple mediators by the fiducial approach are displayed in Figure \ref{fig:mult_med_comparison} for $n=20, 40, 80, 200,$ and $300$. Observing all bars, it is evident that the coverage probability for each indirect effect is either 0.95 or very close to 0.95 for sample sizes of 200 and 300, indicating that the fiducial approach produces robust results for large samples across all five mediators considered for this simulation. The performance for smaller samples (i.e., $n \leq 80$) is weaker in terms of coverage probability, similar to what we have observed in Figure \ref{fig:zi_comparison}, but remains consistent when extended to multiple mediators.

%Similar results are displayed in Figure \ref{fig:di_comparison} for three values of the dispersion parameter: 0.5, 1.0, and 10.0. The fiducial approach shows satisfactory performance in the top left and middle figures; however, for the dispersion parameter of 10.0, to attain the target coverage probability, it requires at least 300 samples for both fiducial and NPB, but the corresponding CW of NPB is much wider compared to that of the fiducial. The delta method remains the worst performer across all combinations of parameters.

%\begin{figure}[H]
    %\centering
    %\includegraphics[width = 1.2\textwidth]{disp_comparison_v2.pdf}
   % \caption{A comparison of  coverage probabilities and widths of confidence intervals  of the fiducial approach, NP bootstrap and delta method  for a single mediator with different dispersion parameter values. The red dashed line represents the 95$\%$ coverage line.}
    %\label{fig:di_comparison}
%\end{figure}

\subsubsection{Sensitivity analysis}
\begin{figure}[H]
    \centering
    \includegraphics[width = 1.1\textwidth, height = 0.30\textheight]{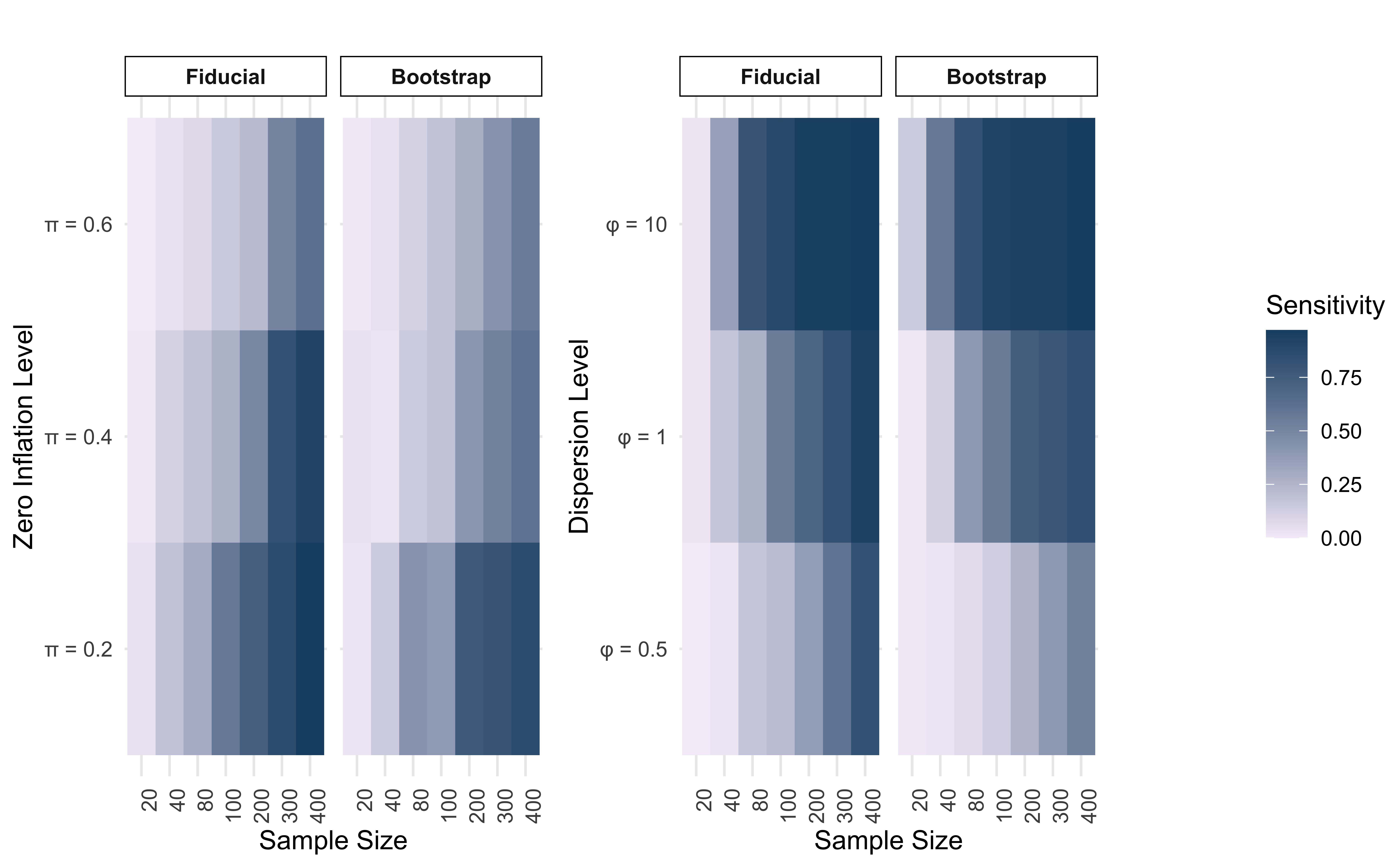}
    \caption{Sensitivity results of fiducial and  NPB  for different zero inflation levels (left panel) and dispersion parameters (right panel) for various sample sizes.} 
    \label{fig:sensitivity_plot}
\end{figure}

In Figure \ref{fig:sensitivity_plot}, it is evident that the fiducial approach shows better sensitivity (i.e. detection rate of non-zero NIEs) compared to NPB, particularly when $\phi = 0.5$, and $\pi = 0.4$ and for other larger values.  Even for larger samples (e.g. 300 and 400)  NPB produced low sensitivity for $\pi=.60$, whereas the fiducial approach demonstrated better sensitivity at the top of the right side of the extreme left cylinder. The sensitivity of the fiducial approach becomes visible starting from $n=80$ and for $\phi=0.5$ at the bottom part of the third cylinder, whereas the NPB fails to show any visible sensitivity even for larger samples (e.g. $n=300$ or $ 400$) and $\pi=0.5$. 
It indicates that the fiducial can better detect NIEs for smaller samples compared to the NPB. Although the delta method is able to detect the NIE  for smaller samples, its corresponding coverage probabilities are poor; hence, they are not shown in Figure \ref{fig:sensitivity_plot}.

\subsection{Rationale behind the findings}

\begin{figure}[H]
    \centering
    \includegraphics[width = 1\textwidth, height = 0.20\textheight]{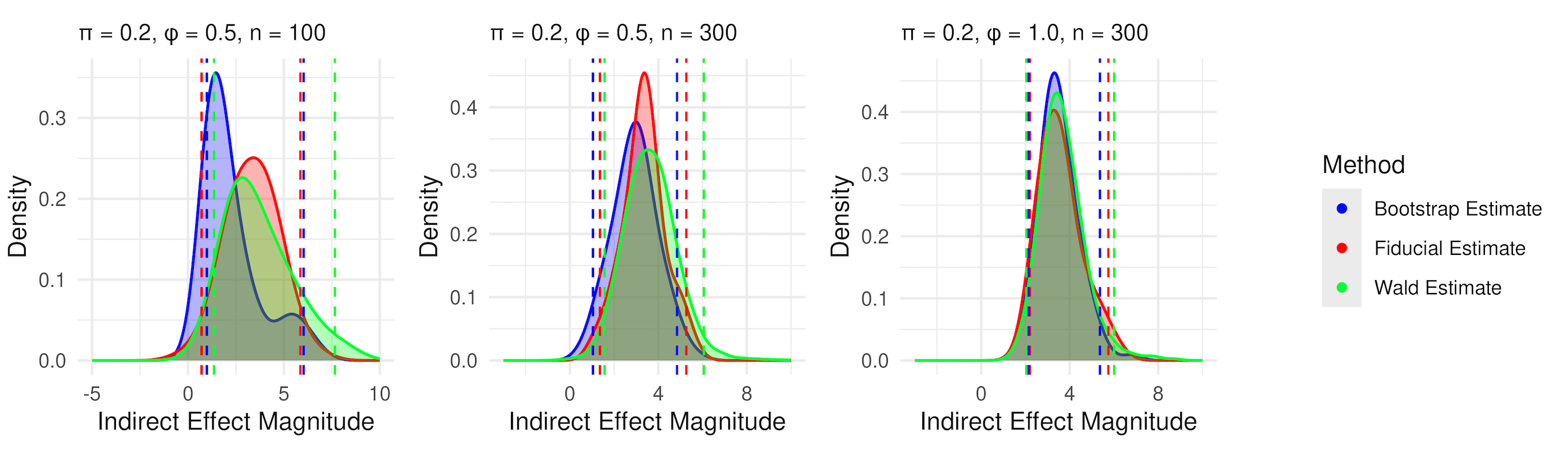} 
    \caption{Comparison of estimated density functions of $\widehat{NIE}$ determined by fiducial, NPB and Wald.}  
    \label{fig:dist_comparison}
\end{figure}

Previous simulations revealed poor performance of the NPB approach for small samples. In Figure \ref{fig:dist_comparison}, estimated density functions of the NIE determined by the fiducial, NPB, and large sample-based Wald approaches are shown. For each scenario in Figure \ref{fig:dist_comparison}, the process was repeated 10,000 times to estimate these density functions. 
Examining the three figures in Figure \ref{fig:dist_comparison},  it is evident that the confidence intervals of NIE determined by the Wald approach (green)  are wider than the corresponding confidence intervals determined by both the fiducial (red) and NPB (blue) approaches. In the left figure, for a small dispersion and sample size, the estimated density function of NPB is right skewed and confidence widths of fiducial and NPB look like the same, but the corresponding coverage probability (not shown in the figure) is much less than $95\%$  compared to that (almost 95$\%$) of the fiducial approach.  All three methods tend to converge for larger dispersion parameters (indicating smaller variances) and a large sample size (i.e.,$n=300$), as seen in the right two figures. The vertical lines represent the lower and upper confidence limits.
\section{Data Analysis}
\label{sec:data_analysis}
This section analyzes the CaCHe Microbiome study data described in  Section \ref{sec:motivation}. Our goal is to investigate whether the vaginal microbiome mediates the relationship between sexual behavior and the risk of BV. We detect pathways identified by the fiducial approach and NPB, adjusted for relevant covariates such as age, socioeconomic status, sexual behavior, WASH (Water, Sanitation, and Hygiene) scores, etc. Additionally, we compare the widths of all confidence intervals constructed by the fiducial and NPB approaches.  The Delta method is excluded in this section due to its poor performance observed in the previous section.
\subsection{Data description}
The CaCHe Microbiome study derives its data from a cohort of 436 schoolgirls in Kenya, providing detailed information on their COVID-related stress, socioeconomic status, WASH  scores, sexual behavior, vaginal microbiome samples, and their respective \textit{NugentBV} scores, where a higher score indicates a high likelihood of acquiring the BV \cite{mehta_analysis_2023}. At baseline, the median age of the subjects was 16.9 years. Among the girls, 33.4\% reported being sexually active, 28.86\% experienced anxiety, and 31.23\% had multiple sexual partners in the past six months. Additionally, 60.07\% had a higher WASH score, and 29.1\% had a higher socioeconomic status.  In our model, the binary exposure variable {\it{sp6mosm}} (A)  indicates whether an individual study subject had multiple sexual partners in the past six months, while the outcome variable (Y) is the continuous \textit{NugentBV} score.
Our primary hypothesis is that engaging in sexual activity with multiple partners increases the risk of acquiring BV, which is partially mediated by changes in the composition of certain taxa. The raw \textit{NugentBV} score was used to estimate both the direct effect of \textit{sp6mosm} on the\textit{ NugentBV} scores and the indirect effect through each taxon. 

\subsubsection{Validity of causal assumptions}
  We verified all key assumptions to ensure the validity of the causal mediation model we used. Causal consistency is upheld because the $Nugent BV$ score of an individual directly reflects her exposure to $sp6mosm$  in the past six months and the corresponding changes in microbiome composition. We minimized the bias by adjusting for age, socioeconomic status, WASH scores, and sexual behavior, while also ensuring that there is no unmeasured confounding in the exposure-mediator or mediator-outcome pathways. Variables related to anxiety and sexual behavior (e.g., anxiety, coerced, transsex) were treated as potential confounders  $(C_2)$  in the exposure-mediator relationship. Similarly, age, socioeconomic status, and WASH score were included as potential confounders  $(C_1)$  in the exposure-outcome relationship, and no confounders were modeled $(C_3)$ for the mediator-outcome relationship. By collecting comprehensive data on all relevant factors and conducting sensitivity analyses, we enhanced the validity of our causal model, enabling us to derive reliable estimates of both the direct and indirect effects. 
  Following \cite{Mehta2020PenileMicrobiome} and a few other researchers (\cite{Ravel2011VaginalMicrobiome}), we primarily focused on the top 10 most abundant taxa, accounting for nearly 70-75 $\%$ of the total raw sequence counts of the microbiome.  The top 10 most abundant taxa in our dataset are: 
\textit{Lactobacillus spp.},
\textit{Lactobacillus iners},
\textit{Lactobacillus crispatus},
\textit{Sneathia sanguinegens}, \textit{Anaerococcus prevotii}, \textit{Gardnerella vaginalis}, \textit{Prevotella timonensis},
\textit{Peptoniphilus timonensis}, 
\textit{Finegoldia magna}, and \textit{Atopobium vaginae}.   
\begin{comment}
  \includegraphics[width=10cm, height=12cm]{HeatmapTxC.png}  
\end{comment}
\subsection{Model selection}
Several models, including mixed-effects Poisson, MEZIPM, and MEZINBM were evaluated for model selection. The Akaike Information Criterion (AIC) identified MEZINBM as the most suitable model for our study data. The smallest AIC value of MEZINBM indicates an optimal balance between model complexity and total number of parameters. The taxon-specific dispersion parameter in the model captures the inherent variability of each taxon. Additionally, the taxon-specific zero-inflation component addresses the excess zeros observed in the raw sequence counts for that taxon, which is a common feature in microbiome data.  Further, to examine how well the MEZINBM fits the study data, we divided the entire range of taxon values into six disjoint cells and calculated the expected number of cell-specific subjects for each taxon, and then used a $\chi^2$ test for goodness of fit. Nine out of ten taxa passed the test, demonstrating our model fits the study data well.
\subsection{Results of the study data}
We constructed GCIs using the fiducial approach and CIs by the NPB for all ten NIEs and determined the corresponding confidence widths. Results are presented in Table \ref{tab:taxa_effects}. Inspecting Table \ref{tab:taxa_effects}, we see that both approaches identified \textit{Lactobacillus crispatus} and \textit{Lactobacillus spp.} as important mediators. An estimate of 0.25 for the NIE in the table indicates that engaging in sexual activity with multiple partners increases the Nugent BV score by 0.25 through changes in the abundance of \textit{Lactobacillus crispatus}. The GCI of \textit{Lactobacillus crispatus} by fiducial is [0.03, 0.47] and the confidence interval by NPB is [0.02, 0.55], indicating that the NIE is positive. Similarly, changes in the abundance of \textit{Lactobacillus spp.} play an important role (NIE =-0.20) in mediating the inverse relationship between multiple sexual partners and the Nugent BV scores. Interestingly, two smaller NIEs were identified by the fiducial approach, but the NPB  failed to detect those two.  Estimate of NIE of \textit{Sneathia sanguinegens} is 0.06, and the corresponding GCI is [0.01, 0.11]. In contrast, the confidence interval by the  NPB has a negative lower bound and a positive upper bound, thus failing to detect its positive relation. 

We observe a similar result for \textit{Peptoniphilus gorbachii}, where the estimate is 0.02, and both bounds of GCI are positive, indicating that it has a non-protective or harmful effect on BV. What is striking the most is the estimate of NIE (-0.14) of \textit{Lactobacillus iners}, though relatively a large number, but inspecting the corresponding GCI and confidence interval, we see that both fiducial and NPB failed to give a clear message. To investigate further, we computed the generalized p-value (.188) using the empirical density function of fiducial approach for testing the non-zero effect of the NIE of \textit{Lactobacillus iners}. Thus, the testing procedure also failed to provide any significant NIE of \textit{Lactobacillus iners}. Both NPB and fiducial are in agreement to detect larger NIEs; however, fiducial produces sharper results for some  very small NIEs.
\begin{table}[htbp]
\centering
\begin{tabular}{|>{\raggedright}m{5cm}|>{\centering}m{3cm}|>{\centering}m{3cm}|>{\centering\arraybackslash}m{3cm}|}
\hline
\textbf{Taxa} & \textbf{NIE estimates} & \textbf{ 95\% GCI by fiducial} & \textbf{95\% CI by NPB} \\
\hline
\textit{Lactobacillus crispatus} & \textbf{0.25}  & \textbf{[0.03, 0.47]} & \textbf{[0.02, 0.55]} \\
\hline
\textit{Lactobacillus iners} & -0.14 & [-0.28, 0.03] & [-0.34, 0.06]\\
\hline
\textit{Lactobacillus spp.} & \textbf{-0.20} & \textbf{[-0.34, -0.04]} & \textbf{[-0.41, -0.01]} \\
\hline
\textit{Gardnerella vaginalis} & 0.03 & [-0.16, 0.23] & [-0.26, 0.38] \\
\hline
\textit{Finegoldia magna} & -0.07 & [-0.37, 0.09] & [-0.54, 0.05] \\
\hline
\textit{Anaerococcus prevotii} & -0.04 & [-0.23, 0.11] & [-0.31, 0.08] \\
\hline
\textit{Atopobium vaginae} & 0.04 & [-0.09, 0.25] & [-0.21, 0.39] \\
\hline
\textit{Sneathia sanguinegens} & \textbf{0.06} & \textbf{[0.01, 0.11]} & [-0.03, 0.10] \\
\hline
\textit{Prevotella timonensis} & -0.04 & [-0.09, 0.04] & [-0.27, 0.13] \\
\hline
\textit{Peptoniphilus gorbachii} & \textbf{0.04} & \textbf{[0.02, 0.11]} & [-0.05, 0.10] \\
\hline
\end{tabular}
\caption{Estimates of NIEs  and corresponding confidence intervals for multiple taxa.}
\label{tab:taxa_effects}
\end{table}

  Inspecting Table 1 we find that \textit{Lactobacillus iners} has a medium effect of -0.14, however, it not statistically significant as corresponding GCI and confidence interval contain zero. To examine whether the NIE of \textit{Lactobacillus iners} is significantly non-zero, we computed the generalized p-value of the corresponding test statistic using the empirical fiducial distribution of the indirect effect, and found it to be 0.188. 

\section{Discussion}
\label{sec:discussion}

Detection of  optimal vaginal microbiome is crucial for developing effective interventions for bacterial vaginosis (BV). Meta-transcriptome studies can help analyze key microbial taxa and their influencing factors. Understanding causal pathways is essential, as BV increases the risk of miscarriage and HIV. Implementing preventive measures, such as limiting sexual partners, can help study changes in microbial abundance. Current treatments include antibiotics, while taxa therapy shows promise but requires further research. Insights from this study may pave the way for new approaches in BV treatment.

This article outlines a detailed procedure for analyzing microbiome data, focusing on identifying those taxa that mediate the relationship between multiple sexual partners  and BV. Confidence interval constructed for each NIE provides a straight  interpretation regarding the significant presence   of the causal effect of the taxon. A joint confidence interval based on an ellipsoid by the multivariate approach fails to provide interpretable results at the individual taxon level, hence that approach is avoided here. The issue of multiple comparison is also avoided as the literature of causal inference of microbiome does not encourage such discussions due to interpretation problems. 

While our approach offers several benefits, it also has some limitations. The Natural Effects model requires the dataset to be replicated for each mediator, which can become complex for high  dimensional  mediators. Moreover, when the sample size is too small, the proposed method may face challenges as it violates the assumption of asymptotic normality for the parameters in the MEZINBM. Our procedure of detection of mediators  is  not developed for high dimensional data. Regularized regression models such as $L_0$   that penalizes the number of non-zero coefficients, $L_1$, (e.g LASSO) or $L_2$, (e.g. Ridge) can be used together with fiducial or NPB for constructing confidence intervals.

Some promising future work involves extending the method to include time-varying exposures, mediators, and outcomes, as multiple time-varying exposures are common in many epidemiological studies. Mediation analysis by grouping taxa  based on the phylogenetic  or Bray-Curtis distance is relevant and can be explored. Additionally, exploring group mediation effects is an area that warrants further investigation.

\bibliographystyle{agsm2}
\bibliography{references4}

\end{document}